\documentclass[aps,prl,twocolumn,showpacs,superscriptaddress,final]{revtex4-1}
\usepackage[final]{graphicx}
\usepackage{amsmath}
\usepackage{epstopdf}
\usepackage{float}
\begin{document}

\title{Inequivalence of Single-Particle and Population Lifetimes in a Cuprate Superconductor}

\author{S.-L. Yang}
\affiliation{Stanford Institute for Materials and Energy Sciences, SLAC National Accelerator Laboratory, 2575 Sand Hill Road, Menlo Park, CA 94025, USA}
\affiliation{Geballe Laboratory for Advanced Materials, Departments of Physics and Applied Physics, Stanford University, Stanford, CA 94305, USA}
\author{J.~A. Sobota}
\affiliation{Stanford Institute for Materials and Energy Sciences, SLAC National Accelerator Laboratory, 2575 Sand Hill Road, Menlo Park, CA 94025, USA}
\affiliation{Advanced Light Source, Lawrence Berkeley National Laboratory, Berkeley, CA 94720, USA}
\author{D. Leuenberger}
\affiliation{Stanford Institute for Materials and Energy Sciences, SLAC National Accelerator Laboratory, 2575 Sand Hill Road, Menlo Park, CA 94025, USA}
\affiliation{Geballe Laboratory for Advanced Materials, Departments of Physics and Applied Physics, Stanford University, Stanford, CA 94305, USA}
\author{Y. He}
\affiliation{Stanford Institute for Materials and Energy Sciences, SLAC National Accelerator Laboratory, 2575 Sand Hill Road, Menlo Park, CA 94025, USA}
\affiliation{Geballe Laboratory for Advanced Materials, Departments of Physics and Applied Physics, Stanford University, Stanford, CA 94305, USA}
\author{M. Hashimoto}
\author{D.~H. Lu}
\affiliation{Stanford Synchrotron Radiation Lightsource, SLAC National Accelerator Laboratory, 2575 Sand Hill Road, Menlo Park, California 94025, USA}
\author{H. Eisaki}
\affiliation{Electronics and Photonics Research Institute, National Institute of Advanced Industrial Science and Technology, Tsukuba, Ibaraki 305-8558, Japan}
\author{P.~S. Kirchmann}
\email{kirchman@slac.stanford.edu}
\affiliation{Stanford Institute for Materials and Energy Sciences, SLAC National Accelerator Laboratory, 2575 Sand Hill Road, Menlo Park, CA 94025, USA}
\author{Z.-X. Shen}
\email{zxshen@stanford.edu}
\affiliation{Stanford Institute for Materials and Energy Sciences, SLAC National Accelerator Laboratory, 2575 Sand Hill Road, Menlo Park, CA 94025, USA}
\affiliation{Geballe Laboratory for Advanced Materials, Departments of Physics and Applied Physics, Stanford University, Stanford, CA 94305, USA}

\date{\today}

\begin{abstract}
We study optimally doped Bi-2212 ($T_\textrm{c}=96$~K) using femtosecond time- and angle-resolved photoelectron spectroscopy. Energy-resolved population lifetimes are extracted and compared with single-particle lifetimes measured by equilibrium photoemission. The population lifetimes deviate from the single-particle lifetimes in the low excitation limit by one to two orders of magnitude. Fundamental considerations of electron scattering unveil that these two lifetimes are in general distinct, yet for systems with only electron-phonon scattering they should converge in the low-temperature, low-fluence limit. The qualitative disparity in our data, even in this limit, suggests that scattering channels beyond electron-phonon interactions play a significant role in the electron dynamics of cuprate superconductors.

\end{abstract}
\pacs{74.72.-h, 78.47.J-, 71.38.-k}
\maketitle

Electron lifetime is a central quantity in condensed matter theories \cite{Ashcroft1976}. It determines macroscopic properties such as electrical and thermal conductivities, and encodes microscopic scattering mechanisms \cite{Ashcroft1976,Mahan2000,Damascelli2003}. Revealing dominant scattering channels in copper-oxide high-temperature superconductors (cuprates) will be key to understanding the complex interplay of orders underlying their phase diagram.

Electron lifetimes in cuprates have been studied in both the energy and time domains. In the energy domain, angle-resolved photoelectron spectroscopy (ARPES) \cite{Devereaux2004, Lanzara2001, Zhou2005, Lee2008, Johnston2012, He2013} and optical spectroscopy \cite{Tu2002, Hwang2004} access the imaginary part of electron self energy $\textrm{Im}\Sigma (\epsilon)$, which is connected to the single-particle lifetime $\tau_\textrm{s}(\epsilon)$ via $\textrm{Im}\Sigma(\epsilon)=\hbar/(2\tau_\textrm{s}(\epsilon))$. This lifetime describes the relaxation process of an excited single particle with energy $\epsilon$. On the other hand, time-resolved reflectivity (trR) \cite{Segre2002,Gedik2004,Kaindl2005} measures a lifetime $\tau_\textrm{p}$ associated with the decay of photoexcited electron population. Systematically comparing $\tau_\textrm{s}(\epsilon)$ and $\tau_\textrm{p}(\epsilon)$ may provide new insights into the underlying scattering mechanisms. To understand the relation between these two lifetimes, one needs to obtain the energy-resolved population lifetime $\tau_\textrm{p}(\epsilon)$ and directly compare with $\tau_\textrm{s}(\epsilon)$. Femtosecond time-resolved ARPES (trARPES) provides this capability \cite{Perfetti2007, Cortes2011, Graf2011, Smallwood2012, Zhang2013, Smallwood2014, Rameau2014, Zhang2014,Sobota2012, Yang2014a}. Several trARPES studies have investigated the relaxation of photo-excited electrons in cuprates \cite{Cortes2011, Graf2011, Smallwood2012, Zhang2013}. Yet, so far, no energy-resolved lifetimes have been extracted from the population dynamics in cuprates. 

In this Letter, we employ trARPES and ARPES with high energy resolution to perform a detailed comparison between $\tau_\textrm{s}(\epsilon)$ and $\tau_\textrm{p}(\epsilon)$ in optimally doped Bi$_\textrm{2}$Sr$_\textrm{2}$Ca$_\textrm{0.92}$Y$_\textrm{0.08}$Cu$_\textrm{2}$O$_{8+\delta}$ (OP Bi-2212, $T_\textrm{c}=96$~K) along the nodal direction \cite{Eisaki2004}. At $20$~K, $\tau_\textrm{p}(\epsilon)$ extracted from trARPES decreases with increasing excitation densities below a characteristic energy of $\sim 60$~meV, yet the trend is reversed above this energy. At first glance, this characteristic energy seems to agree with the mode energies as identified by ARPES measurements of $\tau_\textrm{s}(\epsilon)$, but the absolute values for $\tau_\textrm{s}(\epsilon)$ and $\tau_\textrm{p}(\epsilon)$ are different by $1 \sim 2$ orders of magnitude. This disparity also existed in studies on graphite and graphene \cite{Sugawara2007,Moos2001,Gierz2014}. We demonstrate that $\tau_\textrm{s}(\epsilon)$ and $\tau_\textrm{p}(\epsilon)$ reflect different aspects of electron scattering phenomena and that processes beyond electron-phonon interactions contribute to the disparity. The understanding of this disparity is of importance to future trARPES experiments on all materials.

\begin{figure*}[htp]
\begin{center}
\includegraphics[width=7in]{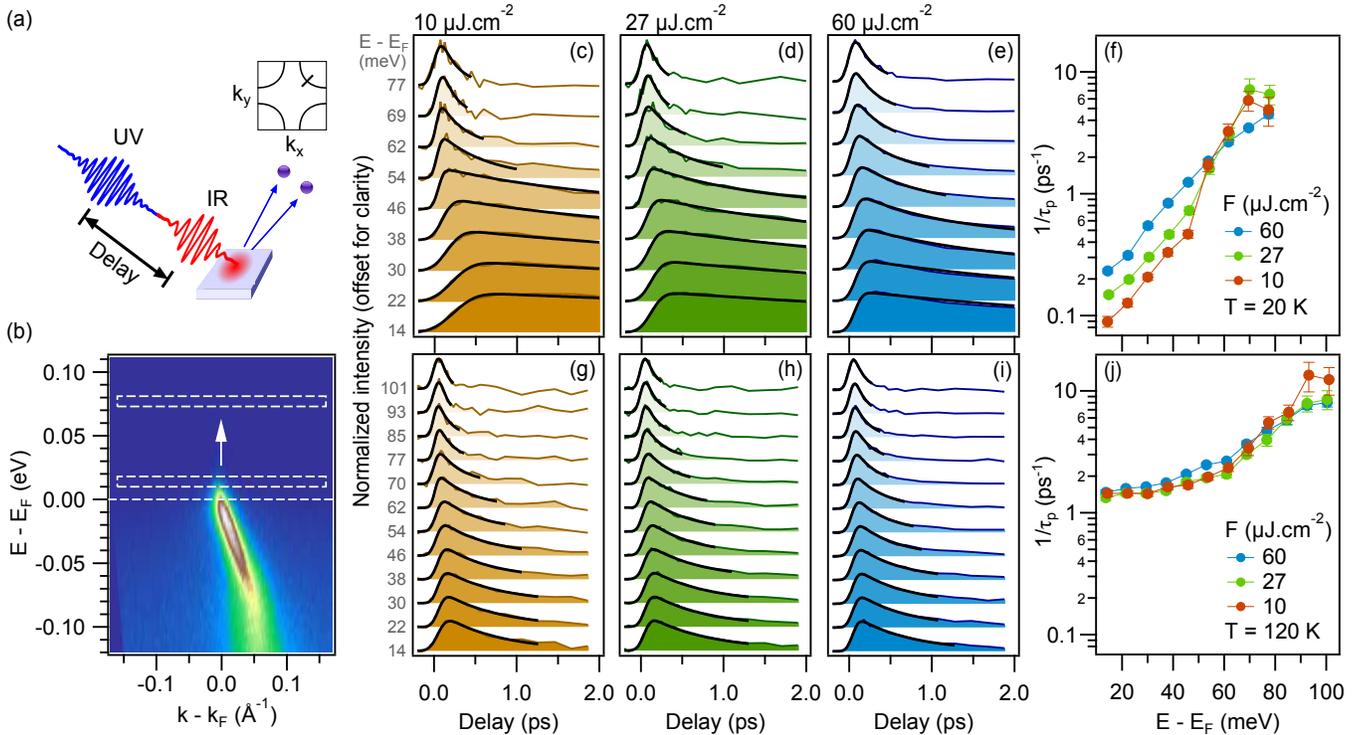}
\caption{Energy-resolved population decay analysis. (a) Scheme of a pump-probe photoemission experiment. (b) Nodal cut obtained by 6 eV photoemission at $20$~K. Boxes mark the windows for momentum integration and 8 meV energy binning used for the population decay analysis. (c)$\sim$(e) Population dynamics normalized by peak intensities at $20$~K for incident excitation densities (fluences) of $10$, $27$, and $60$~$\mu\textrm{J.cm}^{-2}$, respectively. (g)$\sim$(i) Population dynamics at $120$~K for the same set of excitation densities. (f, j) Population decay rates extracted by fitting the transients with a Gaussian-convolved exponential decay at initial delays ($\Delta I(t)>20\%\Delta I_\textrm{max}$). The fitting curves (thick black lines) are overlaid on the population dynamics.}\label{Fig1}
\end{center}
\end{figure*}

Our trARPES setup is based on a Ti-Sapphire regenerative amplifier operating at a repetition rate of $800$~kHz. $1.5$~eV infrared pump pulses excite the sample; $6$~eV ultraviolet probe pulses generate photoelectrons which are collected by a Scienta R4000 analyzer. High quality single crystals of OP Bi-2212~\cite{Eisaki2004} are cleaved in ultrahigh vacuum with a pressure $<7\times10^{-11}$ Torr. Typical energy, momentum, and time resolutions for the trARPES setup are $22$~meV, $0.001$~\AA$^{-1}$, and $100$~fs, respectively. Our ARPES measurement is performed at Beamline 5-4 of Stanford Synchrotron Radiation Lightsource. 7 eV synchrotron light generates photoelectrons which are collected by a Scienta R4000 analyzer. The sample preparation procedure is the same as that for the trARPES measurement. The combined energy resolution is $3$~meV.

We study the cut along the nodal direction in Bi-2212 as shown in Fig.~\ref{Fig1}(b). A clear kink in the band dispersion $\sim70$~meV below the Fermi level $E_\textrm{F}$ can be identified. This indicates a strong bosonic coupling and agrees with previous ARPES investigations \cite{Devereaux2004, Lanzara2001, Zhou2005, Lee2008, Johnston2012, He2013}.

trARPES measurements are also performed on the nodal cut. For the following analysis and discussion, we focus on the photoexcited electron population above $E_\textrm{F}$ \cite{SM}. We integrate over the whole momentum range of the cut as illustrated in Fig.~\ref{Fig1}(b) and subtract the signal before pumping. Energy-resolved transient electron populations are obtained by binning the data into $8$~meV energy intervals and plotting as a function of pump-probe delay. The population dynamics normalized by their peak intensities are shown in Fig.~\ref{Fig1}(c)-(e) for incident fluences of $10$, $27$, and $60$~$\mu\textrm{J.cm}^{-2}$ at $T=20$~K. The population dynamics for the same set of fluences at $120$~K are displayed in Fig.~\ref{Fig1}(g)-(i). These population dynamics are fitted with an exponential decay convolved with a Gaussian function for initial delays as defined by a $20\%$ intensity cutoff \cite{Smallwood2012,Segre2002,SM}. This yields the population decay rates as a function of energy, which are displayed in Fig.~\ref{Fig1}(f) and (j). The fit results are insensitive to the choice of cutoff values \cite{SM}.

We first examine the energy dependence of the population dynamics. A pronounced energy dependence is most clearly observed for the $10$~$\mu\textrm{J.cm}^{-2}$ data at $20$~K (Fig.~\ref{Fig1}(c)). Notably, an abrupt change of the population dynamics occurs near $60$~meV. Below this energy the populations increase for $\sim 0.5$~ps before reaching their maxima, and live as long as a few ps; above this energy the populations reach their maxima near time zero, and decay within a few $100$~fs. Consistently, the corresponding decay rate in Fig.~\ref{Fig1}(f) displays a pronounced increase by one order of magnitude near $60$~meV. The energy dependence for the same fluence at $120$~K is less drastic (Fig.~\ref{Fig1}(g)). Populations at all energies reach their maxima near time zero, and decay in a few $100$~fs. Nevertheless, an abrupt increase near $60\sim 80$~meV is observed in the decay rate (Fig.~\ref{Fig1}(j)). 

\begin{figure}
\begin{center}
\includegraphics[width=\columnwidth]{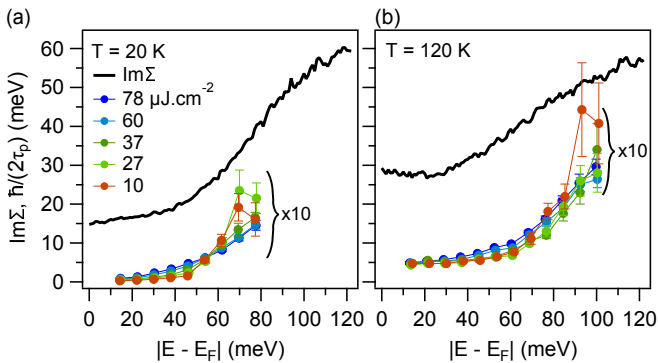}
\caption{Comparison between $\textrm{Im}\Sigma$ (black lines) obtained by ARPES below $E_\textrm{F}$ and population decay rates (solid circles) obtained by trARPES above $E_\textrm{F}$. The method of extracting $\textrm{Im}\Sigma$ from the ARPES data is described in the Supplemental Material~\cite{SM}. Note that the population decay rates are magnified by a factor of 10 for visualization purposes.}\label{Fig2}
\end{center}
\end{figure}

We then study the excitation-density dependence of the population dynamics. The excitation density is characterized by the pump fluence which specifies the incident energy per unit area. At $20$~K, the rising edges become gradually less delayed as the fluence increases. Moreover, the extracted decay rates display a pivoting behavior when tuning the pump fluence (Fig.~\ref{Fig1}(f) and~\ref{Fig2}(a)). Below $60$~meV, the decay rate increases with increasing fluence. Above $60$~meV, the decay rate weakly decreases. This pivoting is much weaker at $120$~K, where the decay rates are approximately fluence independent (Fig.~\ref{Fig1}(j)).

We summarize the three key observations: (i) a $\sim 0.5$~ps rising edge at the lowest fluence and temperature; (ii) an abrupt increase in decay rates near $60\sim 80$~meV at the lowest fluence; (iii) a pivoting behavior for the rate curves when tuning the pump fluence. This complex energy and fluence dependences establish a multi-dimensional constraint for a microscopic understanding. 

A delayed rising edge is usually attributed to cascade processes which fill low-energy states using high-energy electrons \cite{Weida2001,Sobota2012}. For cuprates, the $d$-wave gap indicates that the available low-energy electronic states are only near the node, which provides an additional constraint for electron accumulation. This constraint should be lifted if the superconducting gap is melted either by raising the equilibrium temperature above $T_\textrm{c}$, or by pumping beyond the fluence of $15$~$\mu\textrm{J.cm}^{-2}$ which is required to transiently melt the gap \cite{Zhang2013,Smallwood2014}. Indeed, Fig.~\ref{Fig1} shows that the rising edges of the population dynamics are significantly shortened in both situations.

The abrupt change and the pivoting point of $\tau_\textrm{p}$ occur at the same energy $\sim 60$~meV, which is reminiscent of mode energies revealed by ARPES measurements of $\tau_\textrm{s}$. The pivoting behavior of $\tau_\textrm{p}$ due to electron-phonon coupling has been theoretically predicted \cite{Kemper2014}. This model further predicts that $\tau_\textrm{s}$ and $\tau_\textrm{p}$ converge in the zero excitation limit, yet we find a significant quantitative difference (Fig.~\ref{Fig2}). In particular, if we compare $\hbar/(2\tau_\textrm{p})$ and $\textrm{Im}\Sigma$ below $50$~meV, instead of converging $\hbar/(2\tau_\textrm{p})$ deviates further from $\textrm{Im}\Sigma$ when lowering the fluence. While $\textrm{Im}\Sigma$ extracted from ARPES depends on photon energy, it is always on the same order of magnitude~\cite{Lanzara2001,Johnston2012}. In contrast, the discrepancy between $\tau_\textrm{p}$ from trARPES and $\tau_\textrm{s}$ from ARPES is 1$\sim$2 orders of magnitude. 

While one may relate this discrepancy to unique scattering properties in cuprates, we notice that it is independent of particular material systems. As shown in Table~\ref{Table1}, the discrepancy between $\hbar/(2\tau_\textrm{p})$ and $\textrm{Im}\Sigma$ exists also in graphite and graphene~\cite{Sugawara2007,Moos2001,Gierz2014}.

\begin{table}[ht]
\caption{Comparison of scattering rates at the apparent mode energies ($\Omega_{0}$) obtained by ARPES and trARPES on several materials}
\begin{tabular}{c c c c}
\hline\hline
Material & $\Omega_{0}$ (meV) & $\textrm{Im}\Sigma$ (meV) & $\hbar/(2\tau_\textrm{p})$ (meV) \\
\hline
OP Bi2212 & 70 & 33 & 1.8 \\
Graphite \cite{Sugawara2007,Moos2001} & 200 & 170 & 1.6 \\
Graphene \cite{Gierz2014} & 200 & 118 & 0.37 \\
\hline
\end{tabular}
\label{Table1}
\end{table}

The generality of this discrepancy poses a challenge for connecting time-resolved experiments to single-particle scattering properties. In the following we discuss the conceptual difference between $\tau_\textrm{p}$ and $\tau_\textrm{s}$, and survey representative scattering channels of importance to all materials.

Collision integrals \cite{Ashcroft1976,Groeneveld1995,Gusev1998,Kabanov2008} provide a general formalism for electron scattering processes.

\begin{eqnarray}\label{Eqn1}
\frac{df(\epsilon_{\bf k})}{dt}&=&-\int \frac{d{\bf k'}}{(2\pi)^{3}}W_{\bf k,k'}f(\epsilon_{\bf k})[1-f(\epsilon_{\bf k'})]\nonumber\\
&&+\int \frac{d{\bf k'}}{(2\pi)^{3}}W_{\bf k',k}f(\epsilon_{\bf k'})[1-f(\epsilon_{\bf k})]
\end{eqnarray} 

$\epsilon_{\bf k}$ denotes the electronic state at momentum ${\bf k}$ on a band dispersion. $f(\epsilon_{\bf k})$ is the corresponding occupation. $W_{\bf k,k'}$ stands for the probability of scattering from ${\bf k}$ to ${\bf k'}$. The two integrals in Eqn.~\ref{Eqn1} represent the emptying processes from the state at $\epsilon_{\bf k}$ to other states, and the filling processes from other states back to the state at $\epsilon_{\bf k}$. 

In equilibrium, all the scattering processes reach a detailed balance such that $df(\epsilon_{\bf k})/dt=0$. In non-equilibrium and for perturbative excitations, $f(\epsilon_{\bf k})$ and $f(\epsilon_{\bf k'})$ are transiently changed to $f(\epsilon_{\bf k})+\delta f(\epsilon_{\bf k})$ and $f(\epsilon_{\bf k'})+\delta f(\epsilon_{\bf k'})$. Eqn.~\ref{Eqn1} is expanded to first order in $\delta f$ to describe the evolution of photoexcited electrons~\cite{Kabanov2008,Grimvall1981}. According to the definition of $\tau_\textrm{s}(\epsilon_{\bf k})$, it is associated with excitations that only change $f(\epsilon_{\bf k})$, and hence $\delta f(\epsilon_{\bf k'})=0$ in the Taylor expansion. This occurs in an ARPES measurement where the incident photon probes the same photo-hole as it excites \cite{Damascelli2003}. Using this concept, we derive a general expression for $\tau_\textrm{s}(\epsilon_{\bf k})$.

\begin{eqnarray}\label{Eqn2}
\frac{1}{\tau_\textrm{s}(\epsilon_{\bf k})}&=&-\frac{1}{\delta f(\epsilon_{\bf k})}\frac{d \delta f(\epsilon_{\bf k})}{dt}\nonumber\\
&=&\int \frac{d{\bf k'}}{(2\pi)^{3}}\{W_{\bf k,k'}[1-f(\epsilon_{\bf k'})]+W_{\bf k',k}f(\epsilon_{\bf k'})\}\nonumber\\
\end{eqnarray}

On the other hand, in time-resolved measurements, the pump pulse can excite electrons into many different states independent of the probe pulse. Both $\delta f(\epsilon_{\bf k})$ and $\delta f(\epsilon_{\bf k'})$ are in general non-zero, which leads to Eqn.~\ref{Eqn3} for $\tau_\textrm{p}(\epsilon_{\bf k})$.

\begin{eqnarray}\label{Eqn3}
\frac{1}{\tau_\textrm{p}(\epsilon_{\bf k})} &=& \frac{1}{\tau_\textrm{s}(\epsilon_{\bf k})}-\int \frac{d{\bf k'}}{(2\pi)^{3}}\{W_{\bf k,k'}f(\epsilon_{\bf k})\nonumber\\&&+W_{\bf k',k}[1-f(\epsilon_{\bf k})]\}\frac{\delta f(\epsilon_{\bf k'})}{\delta f(\epsilon_{\bf k})}
\end{eqnarray}

Eqn.~\ref{Eqn3} demonstrates the general distinction between $\tau_\textrm{s}$ and $\tau_\textrm{p}$ without specifying the scattering mechanism. This result is consistent with a number of theoretical investigations which numerically \cite{Groeneveld1995} or analytically \cite{Gusev1998,Kabanov2008} solve the collision integrals. We emphasize that when $\delta f(\epsilon_{\bf k'})$ and $\delta f(\epsilon_{\bf k})$ individually approach zero in the zero excitation limit, their ratio can be nonzero. Therefore, the distinction between $\tau_\textrm{s}$ and $\tau_\textrm{p}$ is fundamental. In the following we illustrate this distinction by surveying a few common scattering channels.

\begin{figure}
\begin{center}
\includegraphics[width=\columnwidth]{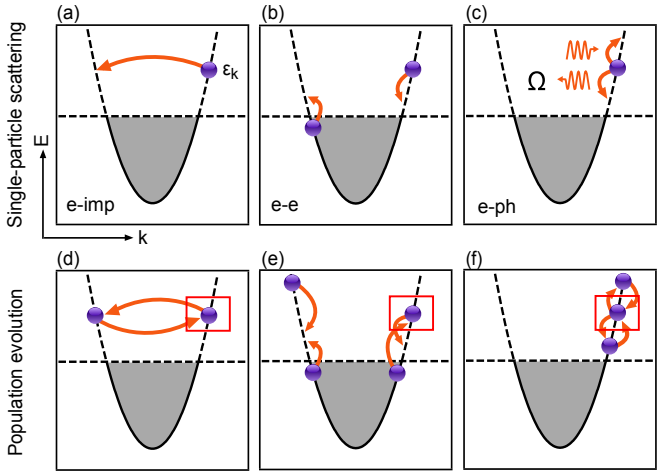}
\caption{Illustration of single-particle scattering and population evolution. (a)$\sim$(c) Single-particle scattering via electron-impurity, electron-electron, and electron-phonon interactions. The scattering rate at $\epsilon_{\bf k}$ arises from the non-equilibrium distribution at $\epsilon_{\bf k}$. (d)$\sim$(f) Population evolution via electron-impurity, electron-electron, and electron-phonon interactions. The population decay rate at $\epsilon_{\bf k}$ arises from non-equilibrium distributions at $\epsilon_{\bf k}$ and other states at $\epsilon_{\bf k'}$. Red boxes denote the integration window as used in our data analysis.}\label{Fig3}
\end{center}
\end{figure}

\emph{Electron-impurity scattering.} This channel is often considered to be elastic and energy-independent \cite{Ashcroft1976}. Since $W_{\bf k, k'}=W_{\bf k', k}$, Eqn.~\ref{Eqn2} and ~\ref{Eqn3} are significantly simplified:

\begin{equation}\label{Eqn4}
\frac{1}{\tau_\textrm{s,e-imp}(\epsilon_{\bf k})} = \int \frac{d{\bf k'}}{(2\pi)^{3}}W_{\bf k,k'}
\end{equation}

\begin{equation}\label{Eqn5}
\frac{1}{\tau_\textrm{p,e-imp}(\epsilon_{\bf k})} = \int \frac{d{\bf k'}}{(2\pi)^{3}}W_{\bf k,k'}[1-\frac{\delta f(\epsilon_{\bf k'})}{\delta f(\epsilon_{\bf k})}]
\end{equation}

Notably, the electron-impurity scattering makes a nonzero contribution to $\textrm{Im}\Sigma$ (Fig.~\ref{Fig3}(a)), but less to $\hbar/(2\tau_\textrm{p})$. The latter contribution completely vanishes for isotropic excitation ($\delta f(\epsilon_{\bf k'})/\delta f(\epsilon_{\bf k})=1$), which results from symmetric scattering between ${\bf k}$ and ${\bf k'}$ (Fig.~\ref{Fig3}(d)). This is consistent with conclusions reached by spherical harmonic decomposition of $f(\epsilon_{\bf k})$~\cite{Goldsman1991,Hennacy1995}. Removing a constant from the experimental $\textrm{Im}\Sigma$ may let us reconcile the discrepancy between $\textrm{Im}\Sigma$ and $\hbar/(2\tau_\textrm{p})$ for low energies (Fig.~\ref{Fig2}). However, the relative increase of $\textrm{Im}\Sigma(\epsilon)$ from $\epsilon=14$~meV to $\epsilon=70$~meV is about $20$~meV, which is still $\sim 10$ times larger than the counterpart for $\hbar/(2\tau_\textrm{p})$.

\emph{Electron-electron scattering.} A photoexcited electron can interact with another electron to redistribute energy and momentum (Fig.~\ref{Fig3}(b)). Time-resolved photoemission experiments on metals have demonstrated that the population lifetimes for electrons $>0.5$~eV above $E_\textrm{F}$ are dominated by electron-electron scattering, and can be related to the single-particle lifetimes \cite{Kirchmann2010,Qui58}. However, for low-energy electrons the filling processes from below $E_\textrm{F}$ become important (Fig.~\ref{Fig3}(e)), which slow down the population decay.

\emph{Electron-phonon scattering.} A photoexcited electron can dissipate its energy by interacting with phonons of energy $\Omega$ (Fig.~\ref{Fig3}(c)). ARPES experiments \cite{Devereaux2004, Lanzara2001, Zhou2005, Lee2008,Johnston2012} have concluded that this is an important scattering channel in cuprates for low-energy electrons $<0.1$~eV. The distinction between $\tau_\textrm{p}$ and $\tau_\textrm{s}$ in Eqn.~\ref{Eqn3} still applies. However, electron-phonon scattering is a special case where this distinction can be mitigated in the low-excitation, low-temperature limit \cite{SM,Sentef2013,Kemper2014}. Scattering processes associated with the photoexcited populations $\delta f(\epsilon_{\bf k'})$ at $\epsilon_{\bf k'}=\epsilon_{\bf k}+\Omega$ and $\epsilon_{\bf k'}=\epsilon_{\bf k}-\Omega$ affect the population at $\epsilon_{\bf k}$ in the form of phonon emission and absorption, respectively. In the low-temperature limit, the phonon absorption processes are negligible due to the vanishing Bose-Einstein distribution. In the low-fluence limit, the phonon emission processes from $\epsilon_{\bf k'}=\epsilon_{\bf k}+\Omega$ to $\epsilon_{\bf k}$ are also negligible since the population at $\epsilon_{\bf k}+\Omega$ can be neglected compared to that at $\epsilon_{\bf k}$.

The three cases above demonstrate that $\tau_\textrm{p}$ is in general distinct from $\tau_\textrm{s}$. Electron-phonon scattering is a special case where $\tau_\textrm{p}$ converges to $\tau_\textrm{s}$ in the low-temperature, low-fluence limit. Notably, when adding more scattering channels the total single-particle scattering rate increases additively. Yet this does not apply to the population decay rate. The main difference, as shown in Eqn.~\ref{Eqn3}, is that certain scattering channels can contribute negatively to the total population decay rate due to filling processes, and slow down the overall population decay. Therefore, although we have a qualitative understanding for each individual scattering channel, understanding their interplay in the population decay is nontrivial. For example, while electron-phonon coupling alone can give rise to a pivoting point of $\tau_\textrm{p}$, the origin of the pivoting point in our data remains unclear.

The discrepancy between $\tau_\textrm{s}$ and $\tau_\textrm{p}$ in our data suggests that scattering channels beyond electron-phonon scattering have an appreciable contribution. Electron-impurity and electron-electron scatterings are likely to contribute. In the superconducting state $W_{\bf k,k'}$ includes pair breaking and reforming processes, and a superconducting gap opens in $\epsilon_{\bf k}$~\cite{Kaplan1976}. The dynamics in the superconducting state can be addressed by the phenomenological Rothwarf-Taylor model, which describes electron pairing and boson-induced pair breaking processes~\cite{Rothwarf1967,Segre2002,Gedik2004,Kaindl2005,Smallwood2012,Zhang2013,Cortes2011,Kabanov2005}. A number of trR~\cite{Segre2002,Gedik2004,Kaindl2005} and trARPES~\cite{Smallwood2012,Zhang2013} experiments have demonstrated that the energy-integrated population lifetimes in cuprates depend on the pump fluence, which is also the case in our data~\cite{SM}. This has been interpreted as dynamics not governed by the boson bottleneck~\cite{Segre2002,Gedik2004,Kaindl2005}. However, due to the nonlinear nature of the Rothwarf-Taylor model, the fluence-dependent decay rates may be present in the bottleneck limit as well~\cite{Kabanov2005,Cortes2011}. To make direct comparison with our energy-resolved rates, a more complete model incorporating the energy-dependence of the pairing processes will be required.

The unique energy- and fluence-dependences of our data serve as a basis for understanding scattering phenomena in cuprates. The delayed rising edges at low temperature reflect the anisotropic gap. The discrepancy between $\tau_\textrm{s}$ and $\tau_\textrm{p}$ suggests scattering channels beyond electron-phonon interactions. Within the formalism we utilize, our data provides a strong constraint for future theories exploring the specific form of $W_{\bf k,k'}$. Our work thus points to a new route to unveil the dominant scattering channels behind high-temperature superconductivity.

\begin{acknowledgments}
{\bf Acknowledgments} We thank Thomas Devereaux, Simon Gerber, Alexander Kemper, Wei-Sheng Lee, Brian Moritz, and Michael Sentef for stimulating discussions. This work was primarily supported by the U.S. Department of Energy, Office of Science, Basic Energy Sciences, Materials Sciences and Engineering Division under contract DE-AC02-76SF00515. S.-L. Y. acknowledges the Stanford Graduate Fellowship. D. L. acknowledges support from the Swiss National Science Foundation, under the Fellowship number P300P2-151328. P. S. K.'s contribution was supported in part by the National Science Foundation under Grant No. PHYS-1066293 and the hospitality of the Aspen Center for Physics.
\end{acknowledgments}

\bibliography{YangSL_Bi2212_scatteringRate_3rdrev_ref}

\begin{thebibliography}{44}
\expandafter\ifx\csname natexlab\endcsname\relax\def\natexlab#1{#1}\fi
\expandafter\ifx\csname bibnamefont\endcsname\relax
  \def\bibnamefont#1{#1}\fi
\expandafter\ifx\csname bibfnamefont\endcsname\relax
  \def\bibfnamefont#1{#1}\fi
\expandafter\ifx\csname citenamefont\endcsname\relax
  \def\citenamefont#1{#1}\fi
\expandafter\ifx\csname url\endcsname\relax
  \def\url#1{\texttt{#1}}\fi
\expandafter\ifx\csname urlprefix\endcsname\relax\def\urlprefix{URL }\fi
\providecommand{\bibinfo}[2]{#2}
\providecommand{\eprint}[2][]{\url{#2}}

\bibitem[{\citenamefont{Ashcroft and Mermin}(1976)}]{Ashcroft1976}
\bibinfo{author}{\bibfnamefont{N.~W.} \bibnamefont{Ashcroft}} \bibnamefont{and}
  \bibinfo{author}{\bibfnamefont{N.~D.} \bibnamefont{Mermin}},
  \emph{\bibinfo{title}{{Solid State Physics}}} (\bibinfo{publisher}{Thomson
  Learning, Inc.}, \bibinfo{year}{1976}).

\bibitem[{\citenamefont{Mahan}(2000)}]{Mahan2000}
\bibinfo{author}{\bibfnamefont{G.~D.} \bibnamefont{Mahan}},
  \emph{\bibinfo{title}{{Many-Particle Physics (Physics of Solids and
  Liquids)}}} (\bibinfo{publisher}{Springer}, \bibinfo{year}{2000}).

\bibitem[{\citenamefont{Damascelli et~al.}(2003)\citenamefont{Damascelli,
  Hussain, and Shen}}]{Damascelli2003}
\bibinfo{author}{\bibfnamefont{A.}~\bibnamefont{Damascelli}},
  \bibinfo{author}{\bibfnamefont{Z.}~\bibnamefont{Hussain}}, \bibnamefont{and}
  \bibinfo{author}{\bibfnamefont{Z.-X.} \bibnamefont{Shen}},
  \bibinfo{journal}{Rev. of Mod. Phys.} \textbf{\bibinfo{volume}{75}},
  \bibinfo{pages}{473} (\bibinfo{year}{2003}).

\bibitem[{\citenamefont{Devereaux et~al.}(2004)\citenamefont{Devereaux, Cuk,
  Shen, and Nagaosa}}]{Devereaux2004}
\bibinfo{author}{\bibfnamefont{T.~P.} \bibnamefont{Devereaux}},
  \bibinfo{author}{\bibfnamefont{T.}~\bibnamefont{Cuk}},
  \bibinfo{author}{\bibfnamefont{Z.-X.} \bibnamefont{Shen}}, \bibnamefont{and}
  \bibinfo{author}{\bibfnamefont{N.}~\bibnamefont{Nagaosa}},
  \bibinfo{journal}{Phys. Rev. Lett.} \textbf{\bibinfo{volume}{93}},
  \bibinfo{pages}{117004} (\bibinfo{year}{2004}).

\bibitem[{\citenamefont{Lanzara~{\it et al.}}(2001)}]{Lanzara2001}
\bibinfo{author}{\bibfnamefont{A.}~\bibnamefont{Lanzara~{\it et al.}}},
  \bibinfo{journal}{Nature} \textbf{\bibinfo{volume}{412}},
  \bibinfo{pages}{510} (\bibinfo{year}{2001}).

\bibitem[{\citenamefont{Zhou~{\it et al.}}(2005)}]{Zhou2005}
\bibinfo{author}{\bibfnamefont{X.~J.} \bibnamefont{Zhou~{\it et al.}}},
  \bibinfo{journal}{Phys. Rev. Lett.} \textbf{\bibinfo{volume}{95}},
  \bibinfo{pages}{117001} (\bibinfo{year}{2005}).

\bibitem[{\citenamefont{Lee~{\it et al.}}(2008)}]{Lee2008}
\bibinfo{author}{\bibfnamefont{W.~S.} \bibnamefont{Lee~{\it et al.}}},
  \bibinfo{journal}{Phys. Rev. B} \textbf{\bibinfo{volume}{77}},
  \bibinfo{pages}{140504 (R)} (\bibinfo{year}{2008}).

\bibitem[{\citenamefont{Johnston~{\it et al.}}(2012)}]{Johnston2012}
\bibinfo{author}{\bibfnamefont{S.}~\bibnamefont{Johnston~{\it et al.}}},
  \bibinfo{journal}{Phys. Rev. Lett.} \textbf{\bibinfo{volume}{108}},
  \bibinfo{pages}{166404} (\bibinfo{year}{2012}).

\bibitem[{\citenamefont{He~{\it et al.}}(2013)}]{He2013}
\bibinfo{author}{\bibfnamefont{J.}~\bibnamefont{He~{\it et al.}}},
  \bibinfo{journal}{Phys. Rev Lett.} \textbf{\bibinfo{volume}{111}},
  \bibinfo{pages}{107005} (\bibinfo{year}{2013}).

\bibitem[{\citenamefont{Tu et~al.}(2002)\citenamefont{Tu, Homes, Gu, Basov, and
  Strongin}}]{Tu2002}
\bibinfo{author}{\bibfnamefont{J.~J.} \bibnamefont{Tu}},
  \bibinfo{author}{\bibfnamefont{C.~C.} \bibnamefont{Homes}},
  \bibinfo{author}{\bibfnamefont{G.~D.} \bibnamefont{Gu}},
  \bibinfo{author}{\bibfnamefont{D.~N.} \bibnamefont{Basov}}, \bibnamefont{and}
  \bibinfo{author}{\bibfnamefont{M.}~\bibnamefont{Strongin}},
  \bibinfo{journal}{Phys. Rev. B} \textbf{\bibinfo{volume}{66}},
  \bibinfo{pages}{144514} (\bibinfo{year}{2002}).

\bibitem[{\citenamefont{Hwang et~al.}(2004)\citenamefont{Hwang, Timusk, and
  Gu}}]{Hwang2004}
\bibinfo{author}{\bibfnamefont{J.}~\bibnamefont{Hwang}},
  \bibinfo{author}{\bibfnamefont{T.}~\bibnamefont{Timusk}}, \bibnamefont{and}
  \bibinfo{author}{\bibfnamefont{G.~D.} \bibnamefont{Gu}},
  \bibinfo{journal}{Nature} \textbf{\bibinfo{volume}{427}},
  \bibinfo{pages}{714} (\bibinfo{year}{2004}).

\bibitem[{\citenamefont{Segre~{\it et al.}}(2002)}]{Segre2002}
\bibinfo{author}{\bibfnamefont{G.~P.} \bibnamefont{Segre~{\it et al.}}},
  \bibinfo{journal}{Phys. Rev. Lett.} \textbf{\bibinfo{volume}{88}},
  \bibinfo{pages}{137001} (\bibinfo{year}{2002}).

\bibitem[{\citenamefont{Gedik~{\it et al.}}(2004)}]{Gedik2004}
\bibinfo{author}{\bibfnamefont{N.}~\bibnamefont{Gedik~{\it et al.}}},
  \bibinfo{journal}{Phys. Rev. B} \textbf{\bibinfo{volume}{70}},
  \bibinfo{pages}{014504} (\bibinfo{year}{2004}).

\bibitem[{\citenamefont{Kaindl et~al.}(2005)\citenamefont{Kaindl, Carnahan,
  Chemla, Oh, and Eckstein}}]{Kaindl2005}
\bibinfo{author}{\bibfnamefont{R.~A.} \bibnamefont{Kaindl}},
  \bibinfo{author}{\bibfnamefont{M.~A.} \bibnamefont{Carnahan}},
  \bibinfo{author}{\bibfnamefont{D.~S.} \bibnamefont{Chemla}},
  \bibinfo{author}{\bibfnamefont{S.}~\bibnamefont{Oh}}, \bibnamefont{and}
  \bibinfo{author}{\bibfnamefont{J.~N.} \bibnamefont{Eckstein}},
  \bibinfo{journal}{Phys. Rev. B} \textbf{\bibinfo{volume}{72}},
  \bibinfo{pages}{060510(R)} (\bibinfo{year}{2005}).

\bibitem[{\citenamefont{Perfetti~{\it et al.}}(2007)}]{Perfetti2007}
\bibinfo{author}{\bibfnamefont{L.}~\bibnamefont{Perfetti~{\it et al.}}},
  \bibinfo{journal}{Phys. Rev. Lett.} \textbf{\bibinfo{volume}{99}},
  \bibinfo{pages}{197001} (\bibinfo{year}{2007}).

\bibitem[{\citenamefont{Cort\'{e}s~{\it et al.}}(2011)}]{Cortes2011}
\bibinfo{author}{\bibfnamefont{R.}~\bibnamefont{Cort\'{e}s~{\it et al.}}},
  \bibinfo{journal}{Phys. Rev. Lett.} \textbf{\bibinfo{volume}{107}},
  \bibinfo{pages}{097002} (\bibinfo{year}{2011}).

\bibitem[{\citenamefont{Graf~{\it et al.}}(2011)}]{Graf2011}
\bibinfo{author}{\bibfnamefont{J.}~\bibnamefont{Graf~{\it et al.}}},
  \bibinfo{journal}{Nature Phys.} \textbf{\bibinfo{volume}{7}},
  \bibinfo{pages}{805} (\bibinfo{year}{2011}).

\bibitem[{\citenamefont{Smallwood~{\it et al.}}(2012)}]{Smallwood2012}
\bibinfo{author}{\bibfnamefont{C.~L.} \bibnamefont{Smallwood~{\it et al.}}},
  \bibinfo{journal}{Science} \textbf{\bibinfo{volume}{336}},
  \bibinfo{pages}{1137} (\bibinfo{year}{2012}).

\bibitem[{\citenamefont{Zhang~{\it et al.}}(2013)}]{Zhang2013}
\bibinfo{author}{\bibfnamefont{W.}~\bibnamefont{Zhang~{\it et al.}}},
  \bibinfo{journal}{Phys. Rev. B} \textbf{\bibinfo{volume}{88}},
  \bibinfo{pages}{245132} (\bibinfo{year}{2013}).

\bibitem[{\citenamefont{Smallwood~{\it et al.}}(2014)}]{Smallwood2014}
\bibinfo{author}{\bibfnamefont{C.~L.} \bibnamefont{Smallwood~{\it et al.}}},
  \bibinfo{journal}{Phys. Rev. B} \textbf{\bibinfo{volume}{89}},
  \bibinfo{pages}{115126} (\bibinfo{year}{2014}).

\bibitem[{\citenamefont{Rameau~{\it et al.}}(2014)}]{Rameau2014}
\bibinfo{author}{\bibfnamefont{J.~D.} \bibnamefont{Rameau~{\it et al.}}},
  \bibinfo{journal}{Phys. Rev. B} \textbf{\bibinfo{volume}{89}},
  \bibinfo{pages}{115115} (\bibinfo{year}{2014}).

\bibitem[{\citenamefont{Zhang~{\it et al.}}(2014)}]{Zhang2014}
\bibinfo{author}{\bibfnamefont{W.}~\bibnamefont{Zhang~{\it et al.}}},
  \bibinfo{journal}{Nature Commun.} \textbf{\bibinfo{volume}{5}},
  \bibinfo{pages}{4959} (\bibinfo{year}{2014}).

\bibitem[{\citenamefont{Sobota~{\it et al.}}(2012)}]{Sobota2012}
\bibinfo{author}{\bibfnamefont{J.~A.} \bibnamefont{Sobota~{\it et al.}}},
  \bibinfo{journal}{Phys. Rev. Lett.} \textbf{\bibinfo{volume}{108}},
  \bibinfo{pages}{117403} (\bibinfo{year}{2012}).

\bibitem[{\citenamefont{Yang~{\it et al.}}(2014)}]{Yang2014a}
\bibinfo{author}{\bibfnamefont{S.-L.} \bibnamefont{Yang~{\it et al.}}},
  \bibinfo{journal}{Appl. Phys. A} \textbf{\bibinfo{volume}{116}},
  \bibinfo{pages}{85} (\bibinfo{year}{2014}).

\bibitem[{\citenamefont{Eisaki~{\it et al.}}(2004)}]{Eisaki2004}
\bibinfo{author}{\bibfnamefont{H.}~\bibnamefont{Eisaki~{\it et al.}}},
  \bibinfo{journal}{Phys. Rev. B} \textbf{\bibinfo{volume}{69}},
  \bibinfo{pages}{064512} (\bibinfo{year}{2004}).

\bibitem[{\citenamefont{Sugawara et~al.}(2007)\citenamefont{Sugawara, Sato,
  Souma, Takahashi, and Suematsu}}]{Sugawara2007}
\bibinfo{author}{\bibfnamefont{K.}~\bibnamefont{Sugawara}},
  \bibinfo{author}{\bibfnamefont{T.}~\bibnamefont{Sato}},
  \bibinfo{author}{\bibfnamefont{S.}~\bibnamefont{Souma}},
  \bibinfo{author}{\bibfnamefont{T.}~\bibnamefont{Takahashi}},
  \bibnamefont{and} \bibinfo{author}{\bibfnamefont{H.}~\bibnamefont{Suematsu}},
  \bibinfo{journal}{Phys. Rev. Lett.} \textbf{\bibinfo{volume}{98}},
  \bibinfo{pages}{036801} (\bibinfo{year}{2007}).

\bibitem[{\citenamefont{Moos et~al.}(2001)\citenamefont{Moos, Gahl, Fasel,
  Wolf, and Hertel}}]{Moos2001}
\bibinfo{author}{\bibfnamefont{G.}~\bibnamefont{Moos}},
  \bibinfo{author}{\bibfnamefont{C.}~\bibnamefont{Gahl}},
  \bibinfo{author}{\bibfnamefont{R.}~\bibnamefont{Fasel}},
  \bibinfo{author}{\bibfnamefont{M.}~\bibnamefont{Wolf}}, \bibnamefont{and}
  \bibinfo{author}{\bibfnamefont{T.}~\bibnamefont{Hertel}},
  \bibinfo{journal}{Phys. Rev Lett.} \textbf{\bibinfo{volume}{87}},
  \bibinfo{pages}{267402} (\bibinfo{year}{2001}).

\bibitem[{\citenamefont{Gierz et~al.}(2014)\citenamefont{Gierz, Link, Starke,
  and Cavalleri}}]{Gierz2014}
\bibinfo{author}{\bibfnamefont{I.}~\bibnamefont{Gierz}},
  \bibinfo{author}{\bibfnamefont{S.}~\bibnamefont{Link}},
  \bibinfo{author}{\bibfnamefont{U.}~\bibnamefont{Starke}}, \bibnamefont{and}
  \bibinfo{author}{\bibfnamefont{A.}~\bibnamefont{Cavalleri}},
  \bibinfo{journal}{Faraday Discussions} \textbf{\bibinfo{volume}{171}},
  \bibinfo{pages}{311} (\bibinfo{year}{2014}).

\bibitem[{SM()}]{SM}
\bibinfo{title}{{See Supplemental Material [url], which includes
  Ref.~\cite{Fried1965}.}}

\bibitem[{\citenamefont{Fried}(1965)}]{Fried1965}
\bibinfo{author}{\bibfnamefont{D.~L.} \bibnamefont{Fried}},
  \bibinfo{journal}{Applied Optics} \textbf{\bibinfo{volume}{4}},
  \bibinfo{pages}{79} (\bibinfo{year}{1965}).

\bibitem[{\citenamefont{Weida et~al.}(2001)\citenamefont{Weida, Ogawa, Nagano,
  and Petek}}]{Weida2001}
\bibinfo{author}{\bibfnamefont{M.~J.} \bibnamefont{Weida}},
  \bibinfo{author}{\bibfnamefont{S.}~\bibnamefont{Ogawa}},
  \bibinfo{author}{\bibfnamefont{H.}~\bibnamefont{Nagano}}, \bibnamefont{and}
  \bibinfo{author}{\bibfnamefont{H.}~\bibnamefont{Petek}}, in
  \emph{\bibinfo{booktitle}{Ultrafast Phenomena XII}} (\bibinfo{year}{2001}),
  vol.~\bibinfo{volume}{66}, p. \bibinfo{pages}{416}.

\bibitem[{\citenamefont{Kemper et~al.}(2014)\citenamefont{Kemper, Sentef,
  Moritz, Freericks, and Devereaux}}]{Kemper2014}
\bibinfo{author}{\bibfnamefont{A.~F.} \bibnamefont{Kemper}},
  \bibinfo{author}{\bibfnamefont{M.~A.} \bibnamefont{Sentef}},
  \bibinfo{author}{\bibfnamefont{B.}~\bibnamefont{Moritz}},
  \bibinfo{author}{\bibfnamefont{J.~K.} \bibnamefont{Freericks}},
  \bibnamefont{and} \bibinfo{author}{\bibfnamefont{T.~P.}
  \bibnamefont{Devereaux}}, \bibinfo{journal}{Phys. Rev. B}
  \textbf{\bibinfo{volume}{90}}, \bibinfo{pages}{075126}
  (\bibinfo{year}{2014}).

\bibitem[{\citenamefont{Groeneveld et~al.}(1995)\citenamefont{Groeneveld,
  Sprik, and Lagendijk}}]{Groeneveld1995}
\bibinfo{author}{\bibfnamefont{R.~H.~M.} \bibnamefont{Groeneveld}},
  \bibinfo{author}{\bibfnamefont{R.}~\bibnamefont{Sprik}}, \bibnamefont{and}
  \bibinfo{author}{\bibfnamefont{A.}~\bibnamefont{Lagendijk}},
  \bibinfo{journal}{Phys. Rev. B} \textbf{\bibinfo{volume}{51}},
  \bibinfo{pages}{11433} (\bibinfo{year}{1995}).

\bibitem[{\citenamefont{Gusev and Wright}(1998)}]{Gusev1998}
\bibinfo{author}{\bibfnamefont{V.~E.} \bibnamefont{Gusev}} \bibnamefont{and}
  \bibinfo{author}{\bibfnamefont{O.~B.} \bibnamefont{Wright}},
  \bibinfo{journal}{Phys. Rev. B} \textbf{\bibinfo{volume}{57}},
  \bibinfo{pages}{2878} (\bibinfo{year}{1998}).

\bibitem[{\citenamefont{Kabanov and Alexandrov}(2008)}]{Kabanov2008}
\bibinfo{author}{\bibfnamefont{V.~V.} \bibnamefont{Kabanov}} \bibnamefont{and}
  \bibinfo{author}{\bibfnamefont{A.~S.} \bibnamefont{Alexandrov}},
  \bibinfo{journal}{Phys. Rev. B} \textbf{\bibinfo{volume}{78}},
  \bibinfo{pages}{174514} (\bibinfo{year}{2008}).

\bibitem[{\citenamefont{Grimvall}(1981)}]{Grimvall1981}
\bibinfo{author}{\bibfnamefont{G.}~\bibnamefont{Grimvall}},
  \emph{\bibinfo{title}{{The Electron-Phonon Interaction in Metals}}}
  (\bibinfo{publisher}{North-Holland Pub. Co.}, \bibinfo{address}{Amsterdam},
  \bibinfo{year}{1981}).

\bibitem[{\citenamefont{Goldsman et~al.}(1991)\citenamefont{Goldsman,
  Henrickson, and Frey}}]{Goldsman1991}
\bibinfo{author}{\bibfnamefont{N.}~\bibnamefont{Goldsman}},
  \bibinfo{author}{\bibfnamefont{L.}~\bibnamefont{Henrickson}},
  \bibnamefont{and} \bibinfo{author}{\bibfnamefont{J.}~\bibnamefont{Frey}},
  \bibinfo{journal}{Solid-State Electron.} \textbf{\bibinfo{volume}{34}},
  \bibinfo{pages}{389} (\bibinfo{year}{1991}).

\bibitem[{\citenamefont{Hennacy et~al.}(1995)\citenamefont{Hennacy, Wu,
  Goldsman, and Mayergoyz}}]{Hennacy1995}
\bibinfo{author}{\bibfnamefont{K.~A.} \bibnamefont{Hennacy}},
  \bibinfo{author}{\bibfnamefont{Y.-J.} \bibnamefont{Wu}},
  \bibinfo{author}{\bibfnamefont{N.}~\bibnamefont{Goldsman}}, \bibnamefont{and}
  \bibinfo{author}{\bibfnamefont{I.~D.} \bibnamefont{Mayergoyz}},
  \bibinfo{journal}{Solid-State Electron.} \textbf{\bibinfo{volume}{38}},
  \bibinfo{pages}{1485} (\bibinfo{year}{1995}).

\bibitem[{\citenamefont{Kirchmann~{\it et al.}}(2010)}]{Kirchmann2010}
\bibinfo{author}{\bibfnamefont{P.~S.} \bibnamefont{Kirchmann~{\it et al.}}},
  \bibinfo{journal}{Nature Phys.} \textbf{\bibinfo{volume}{6}},
  \bibinfo{pages}{782} (\bibinfo{year}{2010}).

\bibitem[{\citenamefont{Quinn and Ferrell}(1958)}]{Qui58}
\bibinfo{author}{\bibfnamefont{J.~J.} \bibnamefont{Quinn}} \bibnamefont{and}
  \bibinfo{author}{\bibfnamefont{R.~A.} \bibnamefont{Ferrell}},
  \bibinfo{journal}{Phys. Rev.} \textbf{\bibinfo{volume}{112}},
  \bibinfo{pages}{812} (\bibinfo{year}{1958}).

\bibitem[{\citenamefont{Sentef~{\it et al.}}(2013)}]{Sentef2013}
\bibinfo{author}{\bibfnamefont{M.}~\bibnamefont{Sentef~{\it et al.}}},
  \bibinfo{journal}{Phys. Rev. X} \textbf{\bibinfo{volume}{3}},
  \bibinfo{pages}{041033} (\bibinfo{year}{2013}).

\bibitem[{\citenamefont{Kaplan~{\it et al.}}(1976)}]{Kaplan1976}
\bibinfo{author}{\bibfnamefont{S.~B.} \bibnamefont{Kaplan~{\it et al.}}},
  \bibinfo{journal}{Phys. Rev. B} \textbf{\bibinfo{volume}{14}},
  \bibinfo{pages}{4854} (\bibinfo{year}{1976}).

\bibitem[{\citenamefont{Rothwarf and Taylor}(1967)}]{Rothwarf1967}
\bibinfo{author}{\bibfnamefont{A.}~\bibnamefont{Rothwarf}} \bibnamefont{and}
  \bibinfo{author}{\bibfnamefont{B.~N.} \bibnamefont{Taylor}},
  \bibinfo{journal}{Phys. Rev. Lett.} \textbf{\bibinfo{volume}{19}},
  \bibinfo{pages}{27} (\bibinfo{year}{1967}).

\bibitem[{\citenamefont{Kabanov et~al.}(2005)\citenamefont{Kabanov, Demsar, and
  Mihailovic}}]{Kabanov2005}
\bibinfo{author}{\bibfnamefont{V.~V.} \bibnamefont{Kabanov}},
  \bibinfo{author}{\bibfnamefont{J.}~\bibnamefont{Demsar}}, \bibnamefont{and}
  \bibinfo{author}{\bibfnamefont{D.}~\bibnamefont{Mihailovic}},
  \bibinfo{journal}{Phys. Rev. Lett.} \textbf{\bibinfo{volume}{95}},
  \bibinfo{pages}{147002} (\bibinfo{year}{2005}).

\end{thebibliography}

\end{document}